\documentclass[lettersize,journal]{IEEEtran}
\usepackage{amsmath,amssymb,bm}
\usepackage{booktabs}
\usepackage{multirow}
\usepackage{graphicx}
\usepackage[justification=centering]{caption}
\usepackage{subcaption}
\usepackage{xcolor}
\usepackage{pgfplots}
\usetikzlibrary{arrows.meta,positioning}
\usepackage{algorithm}
\usepackage{algpseudocode}
\usepackage{cite}
\usepgfplotslibrary{groupplots}
\pgfplotsset{compat=1.18}

\definecolor{figcolor1}{RGB}{227,26,28}
\definecolor{figcolor2}{RGB}{31,120,180}
\definecolor{figcolor3}{RGB}{51,160,44}
\definecolor{figcolor4}{RGB}{255,127,0}
\definecolor{figcolor5}{RGB}{253,191,111}
\definecolor{figcolor6}{RGB}{166,206,227}
\definecolor{figcolor7}{RGB}{178,223,138}
\definecolor{figcolor8}{RGB}{251,154,153}
\definecolor{figcolor9}{RGB}{202,178,214}
\definecolor{figcolor10}{RGB}{106,61,154}
\definecolor{navyblue}{rgb}{0.5, 0.5, 0.5}

\usepackage{hyperref}
\hypersetup{
	colorlinks=true,
	linkcolor=blue,
	filecolor=blue,
	citecolor=blue,
	urlcolor=blue,
	pdftitle={Cost-Aware Neural Early Stopping for LC-OSD},
	pdfpagemode=FullScreen,
}

\newcommand{\EE}{\mathbb{E}}
\newcommand{\ind}{\mathbb{I}}
\newcommand{\abs}[1]{\left|#1\right|}
\DeclareMathOperator*{\argmax}{arg\,max}
\DeclareMathOperator*{\argmin}{arg\,min}

\title{Cost-Aware Neural Early Stopping for \\Local Constraint OSD Decoders}
    \author{\IEEEauthorblockN{Talha Aky{\i}ld{\i}z and Hessam Mahdavifar\\} 	
		\thanks{T. Aky{\i}ld{\i}z is with the EECS Dept., University of Michigan, Ann Arbor, MI, 48104, USA (email: akyildiz@umich.edu).}
		\thanks{H. Mahdavifar is with the EECS Dept., University of Michigan, Ann Arbor, MI, 48104, USA and ECE Dept., Northeastern University, Boston, MA, 02115, USA (email: hessam@umich.edu).}
        \thanks{This work was supported in part by NSF under Grant CCF-2603392.}
        \vspace{-0.8cm}
	}
    
\begin{document}

\maketitle

\begin{abstract}
Local constraint ordered statistics decoding (LC-OSD) provides strong soft decision performance for short block length linear codes, but its practical cost is dominated by the number of tested error patterns (TEPs). This paper proposes a neural early stopping (NES) protocol for LC-OSD with explicit cost control through one trade-off parameter balancing frame error risk and search effort. The proposed approach is trained with frame error rate (FER)-aligned supervision at predefined checkpoints, and learns if additional search is still likely to improve the current best candidate. Later, stopping is decided by comparing predicted continuation need with a cost measured in TEPs. Experimental results across multiple code families show that the proposed protocol significantly reduces average TEP count with only marginal FER degradation, using a single global model for the range of all operating signal-to-noise ratios (SNRs).
\end{abstract}
\vspace{-0.5cm}
\begin{IEEEkeywords}
LC-OSD, ordered statistics decoding, neural early stopping, short block codes.
\end{IEEEkeywords}

\vspace{-0.3cm}
\section{Introduction}
\label{sec:intro}

Ultra-reliable low-latency communications (URLLC) in 5G and beyond impose stringent reliability and latency requirements that necessitate short packets, motivating short blocklength channel codes \cite{chen2018urllc,shirvanimoghaddam2019short}. In this regime, the gap to channel capacity widens, as quantified by finite-blocklength bounds \cite{polyanskiy2010finite}, with maximum-likelihood (ML) decoding being computationally infeasible for general linear codes. Ordered statistics decoding (OSD) \cite{fossorier1995} addresses this with reliability ordering and a search over low-weight test error patterns (TEPs). For a code $\mathcal{C}(n,k)$ of length $n$ and dimension $k$ with minimum distance $d_{\min}$, restricting the search to TEPs of Hamming weight at most $m\approx\lceil d_{\min}/4-1\rceil$ is often sufficient for near-ML performance \cite{yue2021pbosd,yue2025guesswork}. However, the number of TEPs up to weight $m$ is $\sum_{i=0}^{m}\binom{k}{i} = O(k^m)$, which is computationally prohibitive as $k$ grows large. Local constraint OSD (LC-OSD) \cite{wang2022lcosd} mitigates this by filtering candidates through additional parity constraints, though it has a varying complexity which motivates the need for an early stopping mechanism.

Prior work reduces OSD complexity via both reprocessing and preprocessing improvements. Early most reliable basis (MRB) reprocessing refinements considered multiple information sets through iterative information set reduction \cite{fossorier2002isr} and stochastic MRB construction via biased reliabilities \cite{jin2007bias}. More recent methods prune the TEP search and improve the processing order, including probability based OSD with TEP discarding and early stopping \cite{yue2021pbosd}, fast search based OSD that skips higher orders via a predicted cost threshold \cite{choi2019fastscalable}, local constraint based OSD with early termination \cite{wang2022lcosd,liang2023lcosd}, and linear equation OSD that restricts reprocessing to feasible linear systems \cite{yue2022leosd}. Complementary approaches target the preprocessing bottleneck by skipping or simplifying Gaussian elimination using precomputed templates and permuted generator matrices \cite{choi2021fast}, adaptive GE reduction rules \cite{yue2022age}, iterative basis update \cite{li2024ibu}, or RS based parallel candidate generation for BCH codes \cite{yang2022lowlatency}. Other prior work includes efficient early termination rules \cite{wang2021eosd}, and hybrid decoders for short LDPC codes boosted by neural models \cite{li2024boosting}.

In this work, to the best of our knowledge, we propose the first neural early stopping (NES) protocol for LC-OSD, casting early termination as a cost minimization problem. Unlike hand-crafted stopping criteria, the protocol learns from full decoding trajectories and predicts whether continued search is likely to improve the current best codeword at predefined checkpoints. The decoder terminates when the estimated benefit of continuation no longer justifies the additional TEP cost. A tunable scalar parameter selected at deployment traces a continuous complexity--reliability operating curve from a single global model across all SNR points, and experimental results show that the protocol significantly reduces the average TEP count with only marginal FER degradation.

\vspace{-0.3cm}
\section{System Model and Preliminaries}
\subsection{Channel model and soft information}
Consider a binary linear block code $\mathcal{C}(n,k)$ with rate $R=k/n$. Its generator matrix is denoted by $\mathbf{G}\in\mathbb{F}_2^{k\times n}$ and its parity-check matrix by $\mathbf{H}\in\mathbb{F}_2^{(n-k)\times n}$, $\mathbf{G}\mathbf{H}^{\top}=\mathbf{0}$. 
We consider transmission over the canonical binary-input additive white Gaussian noise (B-AWGN) channel. More specifically, each coded bit $c_i\in\{0,1\}$ is mapped to $x_i = 1-2c_i$ for $ i=1,\ldots,n$. The channel output is $\mathbf{y}=[y_1,\ldots,y_n]\in\mathbb{R}^n$ with
$y_i = x_i + n_i$, where $n_i\sim\mathcal{N}(0,\sigma^2)$, for $i=1,\ldots,n$.

The bit-wise hard decision vector is denoted by $\mathbf{z}=[z_1,\ldots,z_n]\in\mathbb{F}_2^n$, obtained by taking the sign of $y_i$'s and mapping them back to the binary field. 
Similarly, the log-likelihood ratio (LLR) vector is denoted by $\boldsymbol{\ell}=[\ell_1,\ldots,\ell_n]\in\mathbb{R}^n$, with
$
    \ell_i = 2y_i / \sigma^2
$, where $\abs{\ell_i}$ is referred to as the reliability of the $i$-th bit.

The ML decoder selects the codeword $\mathbf{v}\in\mathbb{F}_2^n$ satisfying $\mathbf{H}\mathbf{v}^{\top}=\mathbf{0}$ that maximizes the likelihood, i.e.,
\begin{equation}
    \hat{\mathbf{v}}=\argmax_{\mathbf{v}:\mathbf{H}\mathbf{v}^{\top}=\mathbf{0}} p(\mathbf{y} | \mathbf{c}=\mathbf{v}).
\end{equation}

For memoryless channels, including the B-AWGN channel, maximizing the likelihood in the ML decoder reduces to minimizing the soft-weight metric as \vspace{-0.1cm}
\begin{equation}
    \Gamma(\mathbf{v})=\sum_{i=1}^{n} \abs{\ell_i}\ind\left\{v_i\neq z_i\right\},
\label{eq:soft_metric}
\vspace{-0.1cm}
\end{equation}
over all valid codewords. This reliability-weighted discrepancy is the decoding metric used by both OSD and LC-OSD.

\vspace{-0.5cm}
\subsection{Classical OSD}
OSD constructs a systematic generator matrix from the reliability ordering of the received bits and enumerates candidate codewords by perturbing the most reliable positions. Let $\pi$ denote the permutation that sorts bit positions by decreasing reliability, with permuted LLRs $\tilde{\ell}_j=\ell_{\pi(j)}$ satisfying
\begin{equation}
|\tilde{\ell}_1|\ge |\tilde{\ell}_2| \ge \cdots \ge |\tilde{\ell}_n|.
\vspace{-0.15cm}
\end{equation}
The corresponding $n\times n$ permutation matrix is denoted by $\mathbf{\Pi}$. After applying the reliability ordering induced by $\pi$, the ordered generator matrix $\mathbf{G}\mathbf{\Pi}$ is row-reduced to the systematic form $\widetilde{\mathbf{G}}=[\mathbf{I}_k \quad \mathbf{P}]$ where the leading $k$ columns form the most reliable basis (MRB). Let $\hat{\mathbf{u}}_0\in\mathbb{F}_2^k$ be the hard decision vector on the $k$ MRB positions of the permuted received signal.

Candidate codewords are generated by perturbing $\hat{\mathbf{u}}_0$ with a test error pattern $\mathbf{e}\in\mathbb{F}_2^k$ of Hamming weight $w(\mathbf{e})\le m$, i.e.,
$
\hat{\mathbf{c}}(\mathbf{e})=\big[(\hat{\mathbf{u}}_0\oplus\mathbf{e}),(\hat{\mathbf{u}}_0\oplus\mathbf{e})\mathbf{P}\big]\mathbf{\Pi}^{\top}
$, where $m$ denotes the OSD order.
Each candidate is scored by the soft-weight metric as \vspace{-0.1cm}
\begin{equation}
\Gamma(\mathbf{e})=\sum_{i=1}^{n}\abs{\ell_i} \ind\left\{\hat c_i(\mathbf{e})\neq z_i\right\},
\label{eq:gamma_metric}
\vspace{-0.15cm}
\end{equation}
where $\ind\{\cdot\}$ is the indicator function. OSD selects the candidate codeword with the smallest soft-weight metric.

\vspace{-0.45cm}
\subsection{LC-OSD}
\label{ssec:lcosd}
LC-OSD \cite{wang2022lcosd} extends the $k$-position information set of classical OSD to a set of $k+\delta$ most reliable positions, forming an extended MRB, where $\delta\ge 0$ controls the number of additional positions beyond the information set. Denote the index set of these $k+\delta$ positions by $\mathcal{R}$ and the remaining $n-k-\delta$ positions by $\mathcal{L}$. Following the LC-OSD construction, after the reliability-based permutations the parity-check matrix is arranged where the positions in $\mathcal{L}$ appear first and those in $\mathcal{R}$ last, i.e., $\widetilde{\mathbf{H}}=\begin{bmatrix}\mathbf{H}_{\mathcal{L}} & \mathbf{H}_{\mathcal{R}}\end{bmatrix}$. Row-reducing $\widetilde{\mathbf{H}}$ over $\mathbb{F}_2$ yields
\begin{equation}
\widetilde{\mathbf{H}}\sim
\begin{bmatrix}
\mathbf{I}_{n-k-\delta} & \mathbf{P}_1\\
\mathbf{0} & \mathbf{P}_2
\end{bmatrix},
\label{eq:lcosd_hform}
\vspace{-0.1cm}
\end{equation}
where the first $n-k-\delta$ columns correspond to $\mathcal{L}$ and the last $k+\delta$ to $\mathcal{R}$. $\mathbf{P}_1\in\mathbb{F}_2^{(n-k-\delta)\times(k+\delta)}$ is the reconstruction matrix, and $\mathbf{P}_2\in\mathbb{F}_2^{\delta\times(k+\delta)}$ is the local parity-check matrix. The reduced form induces two relations, namely,
\begin{equation}
\tilde{\mathbf{c}}_{\mathcal{L}} = \tilde{\mathbf{c}}_{\mathcal{R}}\mathbf{P}_1^{\top}, \qquad
\tilde{\mathbf{c}}_{\mathcal{R}}\mathbf{P}_2^{\top} = \mathbf{0},
\label{eq:lcosd_parity}
\vspace{-0.15cm}
\end{equation}
where the former reconstructs the positions in $\mathcal{L}$ from $\mathcal{R}$, while the latter imposes $\delta$ local constraints on $\mathcal{R}$ alone.

We denote the hard decision restricted to $\mathcal{R}$ by $\mathbf{z}_{\mathcal{R}}\in\mathbb{F}_2^{k+\delta}$. The matrix $\mathbf{P}_2$ defines a trellis with at most $2^\delta$ states, and a serial list Viterbi algorithm (SLVA) traverses this trellis to generate TEPs $\mathbf{e}_{\mathcal{R}}\in\mathbb{F}_2^{k+\delta}$ satisfying
\begin{equation}
\left(\mathbf{z}_{\mathcal{R}}\oplus\mathbf{e}_{\mathcal{R}}\right)\mathbf{P}_2^{\top}=\mathbf{0}
\quad\Longleftrightarrow\quad
\mathbf{e}_{\mathcal{R}}\mathbf{P}_2^{\top}=\mathbf{z}_{\mathcal{R}}\mathbf{P}_2^{\top},
\label{eq:lcosd_constraint}
\vspace{-0.15cm}
\end{equation}
in non-decreasing order of partial soft-weight metric $\Gamma(\mathbf{e}_{\mathcal{R}})$. For each admissible $\mathbf{e}_{\mathcal{R}}$, the corresponding candidate permuted codeword is reconstructed via
\begin{equation}
\tilde{\mathbf{c}}_{\mathcal{R}}=\mathbf{z}_{\mathcal{R}}\oplus\mathbf{e}_{\mathcal{R}},\qquad
\tilde{\mathbf{c}}_{\mathcal{L}}=\tilde{\mathbf{c}}_{\mathcal{R}}\mathbf{P}_1^{\top},\qquad
\tilde{\mathbf{c}}=[\tilde{\mathbf{c}}_{\mathcal{L}},\,\tilde{\mathbf{c}}_{\mathcal{R}}],
\vspace{-0.15cm}
\end{equation}
and reversing the reliability permutation yields the candidate codeword $\hat{\mathbf{c}}=\tilde{\mathbf{c}}\mathbf{\Pi}^{\top}$. The soft-weight metric is evaluated according to \eqref{eq:soft_metric}, and LC-OSD selects the candidate with the smallest value among all admissible $\mathbf{e}_{\mathcal{R}}$.

Although LC-OSD reduces the search space relative to classical OSD through local constraints, the number of admissible TEPs can still grow prohibitively large as the search proceeds. Consequently, the decoder often expends its full TEP budget even when the correct codeword is found early in the search. This motivates the early stopping protocol, which monitors the decoding trajectory and terminates the search when further exploration is unlikely to improve the current best candidate.

\section{Neural Early Stopping for LC-OSD}
\label{sec:stopping}
\begin{figure*}[t]
\centering
\resizebox{\textwidth}{!}{%
\begin{tikzpicture}[
    box/.style={draw, rounded corners=2pt, minimum height=0.85cm, align=center, font=\scriptsize, inner sep=3pt},
    inputbox/.style={box, fill=blue!10},
    prepbox/.style={box, fill=green!10},
    featbox/.style={box, fill=yellow!12},
    nnbox/.style={box, fill=orange!12},
    decbox/.style={box, fill=red!10},
    outputbox/.style={box, fill=purple!10},
    arrow/.style={-{Latex[length=2mm]}, thick, draw=gray!70},
    node distance=0.3cm
]
    \node[inputbox, text width=1.3cm] (input) {$\boldsymbol{\ell} \in \mathbb{R}^n$\\[1pt]{\tiny Channel LLRs}};

    \node[prepbox, right=0.3cm of input] (prep) {
        {\tiny\textbf{LC-OSD Preprocessing}}\\[1pt]
        {\scriptsize Sort $\pi$, GE $\to \mathcal{L},\mathcal{R},\mathbf{P}_2$}
    };

    \node[prepbox, right=0.3cm of prep] (slva) {
        {\tiny\textbf{SLVA Trellis Search}}\\[1pt]
        {\scriptsize Generate TEPs, update $\hat{\mathbf{c}}_t^\star$}
    };

    \node[featbox, right=0.3cm of slva] (feat) {
        {\tiny\textbf{Checkpoint $t_j$}}\\[1pt]
        {\scriptsize Extract $\phi_j \in \mathbb{R}^{16}$}
    };

    \node[nnbox, right=0.3cm of feat] (mlp) {
        {\tiny\textbf{MLP Estimator $\psi_{\bm{\theta}}$}}\\[1pt]
        {\scriptsize $p_j = \sigma(\psi_{\bm{\theta}}(\phi_j))$}
    };

    \node[decbox, right=0.3cm of mlp] (decision) {
        {\tiny\textbf{Stopping Rule}}\\[1pt]
        {\scriptsize $p_j \le \Delta_j/\lambda$\,?}
    };

    \node[outputbox, right=0.55cm of decision, text width=1.3cm] (output) {$\hat{\mathbf{c}}_{t_j}^\star$\\[1pt]{\tiny Decoded}};

    \draw[arrow] (input) -- (prep);
    \draw[arrow] (prep) -- (slva);
    \draw[arrow] (slva) -- (feat);
    \draw[arrow] (feat) -- (mlp);
    \draw[arrow] (mlp) -- (decision);
    \draw[arrow] (decision) -- node[above, font=\tiny, text=red!70!black] {stop} (output);

    \draw[arrow] (decision.south) -- ++(0,-0.3) -| node[above, pos=0.25, font=\tiny, text=green!50!black] {continue} (slva.south);

\end{tikzpicture}%
}%


\resizebox{\textwidth}{!}{%
\begin{tikzpicture}[
    vecbox/.style={draw, rounded corners=3pt, minimum width=1.6cm, minimum height=1.2cm, align=center, font=\scriptsize},
    scorebox/.style={draw, rounded corners=4pt, minimum width=2.4cm, minimum height=1.0cm, align=center, font=\scriptsize},
    neuron/.style={circle, draw, fill=white, minimum size=0.3cm, inner sep=0pt},
    myarrow/.style={-{Latex[length=2.5mm, width=2mm]}, line width=1pt, draw=black!70},
    node distance=0.4cm
]
    \node[vecbox, fill=yellow!25, text width=3.3cm, minimum height=0.35cm] (fprog) at (0, 0.55) {
        {\tiny\textbf{Progress:} $t_j,\Gamma_{t_j}^\star,\Gamma_{R,t_j}$}
    };

    \node[vecbox, fill=yellow!25, text width=3.3cm, minimum height=0.35cm] (frel) at (0, 0.1) {
        {\tiny\textbf{Reliability:} $\mu,\sigma,m$ over $\mathcal{L},\mathcal{R}$}
    };

    \node[vecbox, fill=yellow!25, text width=3.3cm, minimum height=0.35cm] (fdyn) at (0, -0.35) {
        {\tiny\textbf{Dynamics:} $\Delta\Gamma_j^\star,\Delta\Gamma_{R,j},s_j,u_{t_j}$}
    };

    \node[vecbox, fill=yellow!40, minimum height=0.8cm] (phi) at (4.0, 0.1) {
        {\tiny\textbf{Feature Vector}}\\[2pt]
        $\phi_j \in \mathbb{R}^{16}$
    };

    \coordinate (merge) at (2.8, 0.1);
    \draw[line width=1pt, draw=black!70] (fprog.east) -- ++(0.3,0) |- (merge);
    \draw[line width=1pt, draw=black!70] (frel.east) -- (merge);
    \draw[line width=1pt, draw=black!70] (fdyn.east) -- ++(0.3,0) |- (merge);
    \draw[myarrow] (merge) -- (phi.west);

    \node[neuron, fill=orange!30] (L1a) at (6.5, 0.6) {};
    \node[neuron, fill=orange!30] (L1b) at (6.5, 0.1) {};
    \node[neuron, fill=orange!30] (L1c) at (6.5, -0.4) {};
    \node[font=\tiny, text=gray] at (6.5, -0.85) {16};

    \node[neuron, fill=orange!50] (L2a) at (7.8, 0.75) {};
    \node[neuron, fill=orange!50] (L2b) at (7.8, 0.35) {};
    \node[neuron, fill=orange!50] (L2c) at (7.8, -0.05) {};
    \node[neuron, fill=orange!50] (L2d) at (7.8, -0.45) {};
    \node[font=\tiny, text=gray] at (7.8, -0.85) {128};

    \node[neuron, fill=orange!50] (L3a) at (9.1, 0.75) {};
    \node[neuron, fill=orange!50] (L3b) at (9.1, 0.35) {};
    \node[neuron, fill=orange!50] (L3c) at (9.1, -0.05) {};
    \node[neuron, fill=orange!50] (L3d) at (9.1, -0.45) {};
    \node[font=\tiny, text=gray] at (9.1, -0.85) {128};

    \node[neuron, fill=orange!70] (L4a) at (10.4, 0.15) {};
    \node[font=\tiny, text=gray] at (10.4, -0.85) {1};

    \node[font=\footnotesize\bfseries] at (8.45, 1.3) {MLP $\psi_{\bm{\theta}}$};
    \node[font=\tiny, text=gray] at (8.45, 1.0) {(ReLU, dropout\,=\,0.1)};

    \foreach \i in {L1a,L1b,L1c} \foreach \j in {L2a,L2b,L2c,L2d}
        \draw[gray!40, line width=0.3pt] (\i) -- (\j);

    \foreach \i in {L2a,L2b,L2c,L2d} \foreach \j in {L3a,L3b,L3c,L3d}
        \draw[gray!40, line width=0.3pt] (\i) -- (\j);

    \foreach \i in {L3a,L3b,L3c,L3d}
        \draw[gray!40, line width=0.3pt] (\i) -- (L4a);

    \draw[myarrow] (phi.east) -- (5.9, 0.1);

    \node[vecbox, fill=blue!20, minimum height=0.8cm] (pj) at (12.6, 0.1) {
        \tiny\textbf{Output probability}\\[2pt]
        $p_j = \sigma(o_j)$
    };

    \draw[myarrow] (11.0, 0.1) -- (pj.west);

    \node[draw, rounded corners=4pt, fill=red!12, minimum width=2.6cm, minimum height=1.0cm, align=center, font=\scriptsize] (loss) at (15.8, 0.1) {
        {\tiny\textbf{Training Loss}}\\[2pt]
        $\mathcal{L} = \mathcal{L}_{\mathrm{main}} + \beta\,\mathcal{L}_{\mathrm{mono}}$
    };

    \draw[myarrow] (pj.east) -- (loss.west);

\end{tikzpicture}%
}%
\vspace{-0.2cm}
\caption{Architecture of proposed NES framework for LC-OSD. \textbf{Top}: Runtime pipeline from channel LLRs to decoded codeword. \textbf{Bottom}: MLP estimator with three feature groups and asymmetric training loss.}
\label{fig:pipeline}
\vspace{-0.6cm}
\end{figure*}
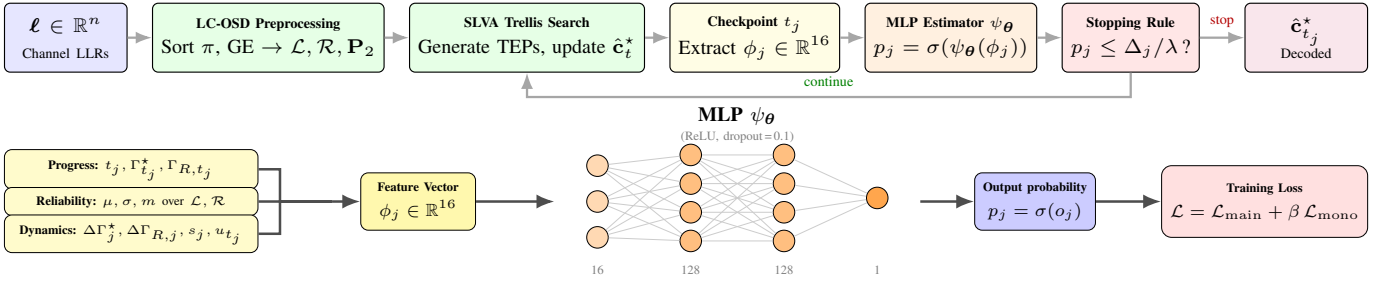

In this section, we present the neural early stopping (NES) framework for LC-OSD. We formulate the checkpointed decision problem where the decoder evaluates a stopping criterion at predetermined points along the search. We then derive a cost-aware stopping rule that balances the risk of premature stopping against the cost of additional TEP evaluations. Next, we define a compact feature representation of the decoder state and a lightweight neural network that estimates whether further search is likely to improve the result. Finally, we present the training objective and the runtime algorithm for deployment, as illustrated in Fig. \ref{fig:pipeline}.

\subsection{Problem Formulation}
\label{ssec:formulation}
For a given frame, the LC-OSD decoder uses the SLVA trellis to enumerate admissible TEPs in non-decreasing partial soft-weight order, up to a maximum budget of $T_{\max}$. Each admissible TEP produces a candidate codeword, and the decoder retains the candidate with the lowest soft-weight metric. Let $T$ denote the number of TEPs examined by the underlying LC-OSD search when no early termination is applied, with each TEP $\mathbf{e}_{\mathcal{R},i}$ producing a candidate codeword $\hat{\mathbf{c}}_i$. The running-best codeword after the first $t$ tested patterns is
\begin{equation}
\hat{\mathbf{c}}_t^\star=\argmin_{1\le i\le t}\Gamma(\hat{\mathbf{c}}_i),
\quad
\Gamma_t^\star=\Gamma(\hat{\mathbf{c}}_t^\star),
\quad
t=1,\ldots,T,
\label{eq:running_best}
\vspace{-0.1cm}
\end{equation}
where the sequence $\{\Gamma_t^\star\}$ is monotone non-increasing in $t$.

Evaluating a stopping criterion after every tested TEP introduces unnecessary overhead. Thus, we query the decision rule only on a predetermined checkpoint grid, i.e.,
\begin{equation}
\mathcal{T}=\{t_1,t_2,\ldots,t_J\},\quad 1\le t_1<\cdots<t_J\le T_{\max},
\label{eq:checkpoints}
\vspace{-0.1cm}
\end{equation}
where $t_j$ is the $j$-th checkpoint and $J$ is the number of queried decision points. At each reached checkpoint, the decoder decides whether to terminate and return the running-best codeword $\hat{\mathbf{c}}_{t_j}^\star$ or continue the LC-OSD search. Let $\tau$ denote an admissible time on this checkpoint grid.

The goal is to minimize the total search effort while penalizing frame errors caused by premature stopping. This is captured by the objective function which is given as 
\begin{equation}
\min_{\tau}\;\EE\left[\tau+\lambda\ind\big\{\hat{\mathbf{c}}_{\tau}^\star\neq \mathbf{c}\big\}\right],
\label{eq:global_obj}
\vspace{-0.1cm}
\end{equation}
where $\mathbf{c}$ is the transmitted codeword and $\lambda>0$ represents a new knob to control the complexity--reliability trade-off, i.e., smaller $\lambda$ favors earlier termination, and larger $\lambda$ places greater emphasis on reliability. Exact optimization of this objective is intractable because the decision at $\tau$ requires knowledge of the remaining search trajectory. Thus, we adopt a local stopping rule to estimate the optimal $\tau$ through sequential per-checkpoint decisions.

\vspace{-0.4cm}
\subsection{Stopping Rule}
\label{ssec:rule}
Our local stopping rule compares the cost of stopping at a given checkpoint $t_j$ with the cost of continuing to the next one. This requires assessing whether continued search beyond $t_j$ will lead to correct decoding. To formalize this, we define the continuation indicator which is given by \vspace{-0.2cm}
\begin{equation}
y_j=
\begin{cases}
  1, & \text{if } \hat{\mathbf{c}}_T^\star=\mathbf{c} \text{ and } \hat{\mathbf{c}}_{t_j}^\star\neq\mathbf{c},\\
0, & \text{otherwise},
\end{cases}
\label{eq:continuation}
\vspace{-0.2cm}
\end{equation}
where $y_j=1$ implies that continued search is necessary to recover the correct codeword, and $y_j=0$ indicates that early termination at $t_j$ would not degrade the decoding outcome.

Since $y_j$ is unknown at decision time, the decoder must estimate $\Pr(y_j=1 | \phi_j)$ from a compact feature vector $\phi_j \in \mathbb{R}^d$ at checkpoint $t_j$. To this end, we employ a lightweight multilayer perceptron (MLP) $\psi_{\bm{\theta}}(\phi_j)\colon \mathbb{R}^d \to \mathbb{R}$ with trainable parameters $\bm{\theta}$. The MLP outputs a logit $o_j = \psi_{\bm{\theta}}(\phi_j)$, and the estimated continuation probability is $p_j = \sigma(o_j)$ where $\sigma$ is the sigmoid function. This quantity approximates the probability that continued search is needed to reach the correct decoding.

The stopping rule compares two costs at each checkpoint. The expected cost of stopping is $C_{\mathrm{stop}}(j) = \lambda p_j$, reflecting the error penalty $\lambda$ weighted by the probability that stopping prevents a correct decoding. The cost of continuing is $C_{\mathrm{cont}}(j) = \Delta_j = t_{j+1} - t_j$, the number of additional TEPs until the next checkpoint. The decoder terminates at the first checkpoint where $C_{\mathrm{stop}}(j) \le C_{\mathrm{cont}}(j)$, equivalently $p_j \le \Delta_j/\lambda$, and sets $\tau = t_j$.

\vspace{-0.4cm}
\subsection{Feature Representation}
\label{ssec:nn}
The raw decoder state at checkpoint $t_j$ is high-dimensional, making it infeasible to be used as a direct input to the MLP. Thus, we construct a compact feature vector $\phi_j \in \mathbb{R}^{16}$ that captures three key aspects of the search: (i) search progress: how far the search has advanced and the quality of the running-best candidate, (ii) channel reliability: the distribution of soft information across the LC-OSD partitions $\mathcal{L}$ and $\mathcal{R}$, and (iii) search dynamics: whether the metric is improving or not.

The progress features include the checkpoint index $t_j$ and the running-best metric $\Gamma_{t_j}^\star$, along with the partial soft weight of the TEP at checkpoint $t_j$ on the searched subset $\mathcal{R}$, $\Gamma_{R,t_j}=\sum_{r\in\mathcal{R}} e_{r,t_j} |\tilde{\ell}_r|$ where $e_{r,t_j}$ denotes the $r$-th entry of $\mathbf{e}_{\mathcal{R},t_j}$.

The reliability features summarize the sorted reliabilities over $\mathcal{L}$ and $\mathcal{R}$ through their mean, standard deviation, and minimum. The mean and standard deviation are
\begin{equation}
\small
\begin{alignedat}{2}
\mu_{L} &= \frac{1}{|\mathcal{L}|}\sum_{i\in\mathcal{L}} |\tilde{\ell}_i|, &\quad
\mu_{R} &= \frac{1}{|\mathcal{R}|}\sum_{i\in\mathcal{R}} |\tilde{\ell}_i|, \\
\sigma_{L} &= \sqrt{\frac{1}{|\mathcal{L}|}\sum_{i\in\mathcal{L}} \big(|\tilde{\ell}_i|-\mu_L\big)^2}, &\quad
\sigma_{R} &= \sqrt{\frac{1}{|\mathcal{R}|}\sum_{i\in\mathcal{R}} \big(|\tilde{\ell}_i|-\mu_R\big)^2},
\end{alignedat}
\end{equation}
and the minima are $m_{L} = \min\limits_{i\in\mathcal{L}} |\tilde{\ell}_i|$ and $m_{R} = \min\limits_{i\in\mathcal{R}} |\tilde{\ell}_i|$.

The search dynamics features track whether the running-best metric continues to improve. The stall age $u_{t_j} = t_j - \max\{t \le t_j : \Gamma_t^\star < \Gamma_{t-1}^\star\}$ counts the number of consecutive TEPs up to $t_j$ without an improvement in $\Gamma_t^\star$. Similarly, $s_j$ counts the number of consecutive checkpoints up to $t_j$ without an improvement in $\Gamma_t^\star$. The metric improvements are $\Delta\Gamma_j^\star = \Gamma_{t_{j-1}}^\star - \Gamma_{t_j}^\star$ and $\Delta\Gamma_{R,j} = \Gamma_{R,t_{j-1}} - \Gamma_{R,t_j}$.

The complete feature vector $\phi_j \in \mathbb{R}^{16}$ concatenates all the above quantities, each normalized to ensure comparable magnitudes across features, and is given by
\begingroup
\small
\begin{align}
\phi_j = \bigg[&\frac{\log_2 t_j}{\log_2 T_{\max}}, \frac{\Gamma_{t_j}^{\star}}{S}, \frac{\Gamma_{R,t_j}}{S}, \frac{\Gamma_{t_j}^{\star}-\Gamma_{R,t_j}}{S}, \notag \\[4pt]
&\frac{\mu_L}{\bar a}, \frac{\sigma_L}{\bar a}, \frac{m_L}{\bar a}, \frac{\mu_R}{\bar a}, \frac{\sigma_R}{\bar a}, \frac{m_R}{\bar a}, \frac{\delta}{n-k}, \frac{|\mathcal{L}|}{n-k}, \notag\\[4pt]
&\frac{\Delta\Gamma_j^{\star}}{S}, \frac{\Delta\Gamma_{R,j}}{S}, \min \bigg(1,\frac{s_j}{s_{\mathrm{sat}}}\bigg), \frac{\log_2\max\{1,u_{t_j}\}}{\log_2 T_{\max}}\bigg]^\top,
\label{eq:features}
\end{align}
\endgroup
where $S=\sum_{i=1}^{n}\abs{\ell_i}$, $\bar a=S/n$, and $s_{\mathrm{sat}}=32$. The total soft weight $S$ and average reliability $\bar{a}$ serve as normalizing constants that make the features invariant to the absolute scale of the channel LLRs, enabling the NES to generalize across $E_b/N_0$ operating points. These three feature groups collectively summarize the decoder state at each checkpoint.

\vspace{-0.15cm}
\subsection{Training Objective}
\label{ssec:training}
The proposed approach is trained offline from full-budget decoding trajectories, where every frame is decoded to completion without early stopping. Each visited checkpoint in such a trajectory contributes one training sample consisting of the feature vector $\phi_j$ and the continuation label $y_j$ defined in \eqref{eq:continuation}.

The goal of training is to produce an estimator that accurately predicts the continuation probability $p_j$. Since premature stopping can cause a decoding error while unnecessary continuation only wastes computation, the loss must penalize the former more heavily. To reflect this asymmetry, the primary loss for a frame with $J$ visited checkpoints is given as \vspace{-0.15cm}
\begin{equation}
\mathcal{L}_{\mathrm{main}}\!=\!
\frac{1}{J}\!\sum_{j=1}^{J}\!
\Big[\!\!
\underbrace{\alpha y_{j} \ell_{\mathrm{sp}}(-o_{j})}_{\text{premature-stop penalty}}
\!\!+\!\!\!\!\underbrace{(1 - y_{j})\,\frac{r_{j}}{\kappa} \ell_{\mathrm{sp}}(o_{j})}_{\text{unnecessary-continuation penalty}}
\!\!\!\!\Big],
\vspace{-0.15cm}
\end{equation}
where $\ell_{\mathrm{sp}}(x)=\log(1+e^x)$ is the softplus function, $r_j = T - t_j$ is the number of TEPs remaining after checkpoint $t_j$, $\alpha>0$ weights premature stopping, and $\kappa>0$ normalizes the residual-effort term. The first term penalizes stopping when search is still needed to reach the correct codeword ($y_j = 1$), while the second penalizes unnecessary continuation ($y_j = 0$) with the cost scaled by $r_j$ to reflect the remaining search effort.

In addition, as the search progresses, the continuation probability should monotonically decrease. Hence, we add a regularizer given by \vspace{-0.2cm}
\begin{equation}
\mathcal{L}_{\mathrm{mono}}=
\frac{1}{J-1}\sum_{j=1}^{J-1}\operatorname{ReLU}(p_{j+1}-p_j),
\vspace{-0.1cm}
\end{equation}
where $\operatorname{ReLU}(x)=\max(0,x)$. Combining the primary loss with the monotonicity regularizer, the full training objective is $\mathcal{L}=\mathcal{L}_{\mathrm{main}}+\beta\mathcal{L}_{\mathrm{mono}}$ with $\beta>0$ controlling the monotonicity penalty strength. The MLP consists of two hidden layers with 128 neurons each, ReLU activations, and a dropout rate of 0.1, and is trained by minimizing $\mathcal{L}$ over mini-batches of randomly selected frames. The resulting runtime deployment procedure is summarized in Algorithm \ref{alg:runtime}.

\setlength{\textfloatsep}{1pt}
\begin{algorithm}[t]
\small
\caption{Runtime NES for LC-OSD}
\label{alg:runtime}
\begin{algorithmic}[1]
\State \textbf{Input:} LLR vector $\boldsymbol{\ell}$, checkpoint set $\{t_1,\ldots,t_J\}$, trained estimator $\psi_{\bm{\theta}}$, control parameter $\lambda$, budget $T_{\max}$
\State \textbf{Initialize:} running-best metric $\Gamma^\star\leftarrow +\infty$, running-best codeword $\hat{\mathbf{c}}^\star\leftarrow \varnothing$, checkpoint index $j\leftarrow 1$, TEP index $t\leftarrow 1$
\State Define $t_{J+1}\leftarrow T_{\max}$
\While{$t\le T_{\max}$ \textbf{and} LC-OSD search is not exhausted}
  \State Execute one LC-OSD expansion step at index $t$
  \State Update $\hat{\mathbf{c}}_t^\star$ and $\Gamma_t^\star$ if improved
  \If{$j\le J$ \textbf{and} $t=t_j$}
      \State Build normalized feature vector $\phi_j$ 
      \State Compute continuation probability $p_j=\sigma \left(\psi_{\bm{\theta}}(\phi_j)\right)$
      \State Compute $\Delta_j=t_{j+1}-t_j$
      \If{$p_j \le \Delta_j/\lambda$}
          \State \textbf{return} $\hat{\mathbf{c}}_{t_j}^\star$ \Comment{early termination}
      \EndIf
      \State $j\leftarrow j+1$
  \EndIf
  \State $t\leftarrow t+1$
\EndWhile
\State \textbf{return} $\hat{\mathbf{c}}_T^\star$ \Comment{search exhausted or budget reached}
\end{algorithmic}
\end{algorithm}

\vspace{-0.3cm}
\section{Numerical Results}
\label{sec:results}
This section presents simulation results to demonstrate the effectiveness of the NES protocol for LC-OSD decoding. Three code families are considered: extended BCH (eBCH), Reed--Muller (RM), and LDPC codes. All simulations assume the B-AWGN channel. For each code, we report two complementary performance metrics: the frame error rate (FER), and the average number of TEPs per frame.

\begin{figure}[t]
\centering
\begin{subfigure}{\columnwidth}
\centering
\begin{tikzpicture}
\begin{axis}[
    name=ebchfer,
    width=\columnwidth,
    height=0.6\columnwidth,
    xmin=0, xmax=3,
    xtick={0,0.5,1,1.5,2,2.5,3},
    ymode=log,
    ymin=1e-5, ymax=1,
    xlabel={$E_b/N_0$ (dB)},
    ylabel={FER},
    grid=both,
    grid style={dotted,gray!50},
    tick label style={font=\scriptsize},
    label style={font=\small},
    legend style={font=\tiny, draw=none, fill=none, at={(0.02,0)}, anchor=south west},
    legend columns=1,
]
\addplot+[figcolor1, solid, line width=0.7pt, mark=o, mark size=1.8pt, mark options={solid, fill=white}] coordinates {(0,0.47961) (0.5,0.25967) (1,0.13095) (1.5,0.049992) (2,0.01341) (2.5,0.00292) (3,0.000451) (3.5,0.000058)};
\addlegendentry{NES, $\lambda=384$}
\addplot+[figcolor2, solid, line width=0.7pt, mark=diamond*, mark size=1.9pt, mark options={solid, fill=figcolor2}] coordinates {(0,0.46948) (0.5,0.24557) (1,0.11718) (1.5,0.04199) (2,0.010699) (2.5,0.00204) (3,0.000286) (3.5,0.000026)};
\addlegendentry{NES, $\lambda=1024$}
\addplot+[figcolor3, solid, line width=0.7pt, mark=triangle, mark size=1.9pt, mark options={solid, fill=white}] coordinates {(0,0.46147) (0.5,0.23889) (1,0.11265) (1.5,0.037658) (2,0.0094437) (2.5,0.0016677) (3,0.000212) (3.5,0.000023)};
\addlegendentry{NES, $\lambda=2048$}
\addplot+[figcolor10, solid, line width=0.7pt, mark=square*, mark size=1.8pt, mark options={solid, fill=figcolor10}] coordinates {(0,0.47733) (0.5,0.25157) (1,0.11007) (1.5,0.03553) (2,0.007596) (2.5,0.001144) (3,0.000102) (3.5,0.000002)};
\addlegendentry{DAI}
\addplot+[figcolor9, solid, line width=0.7pt, mark=triangle*, mark size=1.9pt, mark options={solid, fill=figcolor9}] coordinates {(0,0.47733) (0.5,0.25157) (1,0.11007) (1.5,0.03553) (2,0.007596) (2.5,0.001144) (3,0.000102) (3.5,0.000002)};
\addlegendentry{SAI}
\addplot+[gray, solid, line width=0.7pt, mark=star, mark size=1.9pt, mark options={solid, fill=gray}] coordinates {(0,0.44385) (0.5,0.22852) (1,0.10272) (1.5,0.03363) (2,0.00742) (2.5,0.00110) (3,0.00007) (3.5,0.00002)};
\addlegendentry{TSC}
\addplot+[figcolor6, solid, line width=0.7pt, mark=pentagon*, mark size=1.9pt, mark options={solid, fill=figcolor6}] coordinates {(0,0.32543) (0.5,0.24244) (1,0.16081) (1.5,0.10439) (2,0.05598) (2.5,0.02950) (3,0.01386) (3.5,0.00586)};
\addlegendentry{NES, $\lambda=12$}
\addplot+[figcolor10, solid, line width=0.7pt, mark=square*, mark size=1.8pt, mark options={solid, fill=figcolor10}, forget plot] coordinates {(0,0.32938) (0.5,0.24588) (1,0.16305) (1.5,0.10601) (2,0.05590) (2.5,0.02891) (3,0.01347) (3.5,0.00555)};
\addplot+[figcolor9, solid, line width=0.7pt, mark=triangle*, mark size=1.9pt, mark options={solid, fill=figcolor9}, forget plot] coordinates {(0,0.32938) (0.5,0.24588) (1,0.16194) (1.5,0.10481) (2,0.05593) (2.5,0.02868) (3,0.01356) (3.5,0.00555)};
\addplot+[gray, solid, line width=0.7pt, mark=star, mark size=1.9pt, mark options={solid, fill=gray}, forget plot] coordinates {(0,0.32268) (0.5,0.23759) (1,0.15783) (1.5,0.10483) (2,0.05338) (2.5,0.02880) (3,0.01329) (3.5,0.00548) (4,0.00178)};
\node[font=\scriptsize\bfseries\sffamily, color=navyblue] at (axis cs:2.8,0.003) {\textsf{\textbf{n\,=\,128}}};
\node[font=\scriptsize\bfseries\sffamily, color=navyblue] at (axis cs:2.8,0.035) {\textsf{\textbf{n\,=\,32}}};
\draw[black, thin] (axis cs:1.9,0.005) rectangle (axis cs:2.1,0.02);
\coordinate (ebchzoom) at (axis cs:2.1,0.005);
\end{axis}
\begin{axis}[
    name=ebchinset,
    at={(ebchfer.south east)}, anchor=south east,
    xshift=-50pt, yshift=15pt,
    width=0.32\columnwidth, height=0.32\columnwidth,
    xmin=1.9, xmax=2.1, xtick={2},
    ymode=log, ymin=0.005, ymax=0.02,
    ytick={0.005, 0.01, 0.02},
    yticklabels={$5\!\times\!10^{-3}$, $10^{-2}$, $2\!\times\!10^{-2}$},
    grid=both, grid style={dotted,gray!50},
    tick label style={font=\tiny},
    axis background/.style={fill=white},
]
\addplot+[figcolor1, solid, line width=0.7pt, mark=o, mark size=1.4pt, mark options={solid, fill=white}] coordinates {(1.5,0.049992) (2,0.01341) (2.5,0.00292)};
\addplot+[figcolor2, solid, line width=0.7pt, mark=diamond*, mark size=1.4pt, mark options={solid, fill=figcolor2}] coordinates {(1.5,0.04199) (2,0.010699) (2.5,0.00204)};
\addplot+[figcolor3, solid, line width=0.7pt, mark=triangle, mark size=1.4pt, mark options={solid, fill=white}] coordinates {(1.5,0.037658) (2,0.0094437) (2.5,0.0016677)};
\addplot+[figcolor10, solid, line width=0.7pt, mark=square*, mark size=1.4pt, mark options={solid, fill=figcolor10}] coordinates {(1.5,0.03553) (2,0.007596) (2.5,0.001144)};
\addplot+[figcolor9, solid, line width=0.7pt, mark=triangle*, mark size=1.4pt, mark options={solid, fill=figcolor9}] coordinates {(1.5,0.03553) (2,0.007596) (2.5,0.001144)};
\addplot+[gray, solid, line width=0.7pt, mark=star, mark size=1.4pt, mark options={solid, fill=gray}] coordinates {(1.5,0.03363) (2,0.00742) (2.5,0.00110)};
\end{axis}
\draw[->, black, thin] (ebchzoom) -- (ebchinset.north);
\end{tikzpicture}%
\vspace{-0.3cm}
\caption{Frame error rate.}
\label{fig:ebch-fer}
\end{subfigure}\\[0.0cm]
\begin{subfigure}{\columnwidth}
\centering
\begin{tikzpicture}
\begin{axis}[
    width=\columnwidth,
    height=0.6\columnwidth,
    xmin=0, xmax=3,
    xtick={0,0.5,1,1.5,2,2.5,3},
    ymode=log,
    ymin=1, ymax=20000,
    xlabel={$E_b/N_0$ (dB)},
    ylabel={Average TEP},
    grid=both,
    grid style={dotted,gray!50},
    tick label style={font=\scriptsize},
    label style={font=\small},
    legend style={font=\tiny, draw=none, fill=none, at={(0.02,0)}, anchor=south west},
    legend columns=1,
]
\addplot+[figcolor1, solid, line width=0.7pt, mark=o, mark size=1.8pt, mark options={solid, fill=white}] coordinates {(0,407.0) (0.5,269.0) (1,152.7) (1.5,69.8) (2,25.4) (2.5,7.8) (3,2.6) (3.5,1.3)};
\addlegendentry{NES, $\lambda=384$}
\addplot+[figcolor2, solid, line width=0.7pt, mark=diamond*, mark size=1.9pt, mark options={solid, fill=figcolor2}] coordinates {(0,825.2) (0.5,519.3) (1,283.2) (1.5,119.7) (2,40.5) (2.5,11.0) (3,2.9) (3.5,1.4)};
\addlegendentry{NES, $\lambda=1024$}
\addplot+[figcolor3, solid, line width=0.7pt, mark=triangle, mark size=1.9pt, mark options={solid, fill=white}] coordinates {(0,1310.9) (0.5,817.7) (1,435.3) (1.5,181.9) (2,58.0) (2.5,14.4) (3,3.3) (3.5,1.4)};
\addlegendentry{NES, $\lambda=2048$}
\addplot+[figcolor10, solid, line width=0.7pt, mark=square*, mark size=1.8pt, mark options={solid, fill=figcolor10}] coordinates {(0,524.6) (0.5,588.4) (1,501.7) (1.5,339.3) (2,165.1) (2.5,58.7) (3,17.1) (3.5,5.3)};
\addlegendentry{DAI}
\addplot+[figcolor9, solid, line width=0.7pt, mark=triangle*, mark size=1.9pt, mark options={solid, fill=figcolor9}] coordinates {(0,524.6) (0.5,588.4) (1,501.7) (1.5,339.3) (2,165.0) (2.5,58.7) (3,17.1) (3.5,5.3)};
\addlegendentry{SAI}
\addplot+[gray, solid, line width=0.7pt, mark=star, mark size=1.9pt, mark options={solid, fill=gray}] coordinates {(0,15045.1) (0.5,13222.0) (1,10516.2) (1.5,7327.2) (2,4211.0) (2.5,1905.1) (3,676.0) (3.5,170.8)};
\addlegendentry{TSC}
\end{axis}
\end{tikzpicture}%
\vspace{-0.3cm}
\caption{Average tested error patterns.}
\label{fig:ebch-tep}
\end{subfigure}
\caption{FER and average TEP versus $E_b/N_0$ for $\mathcal{C}_1[128,64]$ and $\mathcal{C}_1[32,16]$.}
\vspace{-0.1cm}
\label{fig:ebch}
\end{figure}
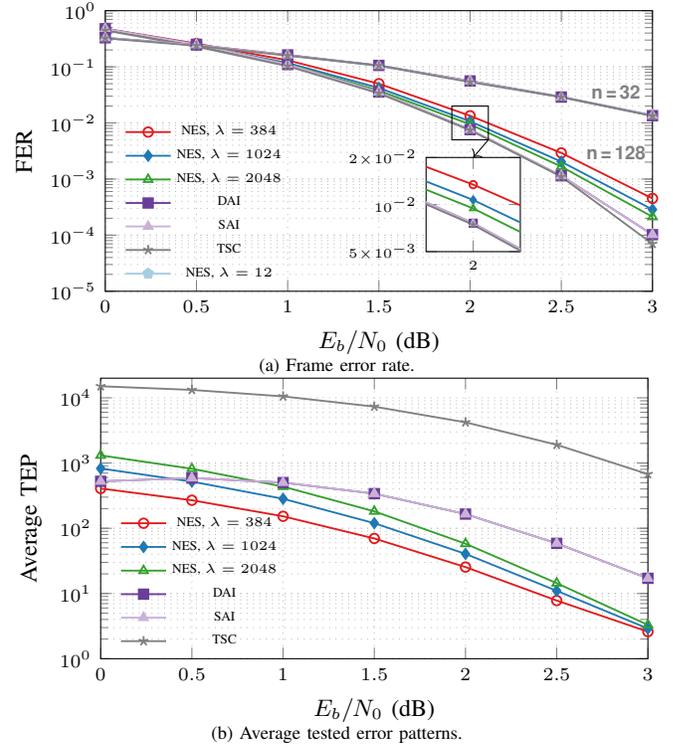

The NES protocol $\psi_{\bm{\theta}}$ is trained offline on full-budget LC-OSD decoding trajectories with $\delta=8$ and $T_{\max}=2^{14}$. Training data is collected from $10^5$ randomly generated frames per code. The stopping criterion is evaluated at a predetermined set of checkpoints from $t_1=1$ to $t_J=T_{\max}$, with denser spacing at low TEP counts. The training loss hyperparameters are set to $\alpha=12$, $\kappa=T_{\max}$, and $\beta=0.05$. The model is optimized for $12{,}000$ steps using Adam optimizer with learning rate $5\times10^{-4}$, weight decay $10^{-4}$, and gradient clipping at $1.0$. In testing, a single trained model is used across all SNRs (in terms of $E_b/N_0$), and the control knob $\lambda$ is varied to obtain different operating points. The proposed protocol is compared with the conventional dynamic approximate ideal (DAI), static approximate ideal (SAI), and trivial stopping criterion (TSC) described in \cite{liang2023lcosd}, which serve as baselines with progressively more conservative early termination.

Fig. \ref{fig:ebch} illustrates the FER and average TEP count for eBCH codes $\mathcal{C}_1[128,64]$ and $\mathcal{C}_1[32,16]$ across different $E_b/N_0$ values using $10^6$ frames per point. The proposed protocol is evaluated at three operating points $\lambda = 384$, $1024$, and $2048$. At the most aggressive setting ($\lambda=384$), the average number of TEPs is reduced by a factor of $3\text{--}8\times$ relative to DAI and by a substantially larger factor relative to TSC across the mid-to-high $E_b/N_0$ range, at the cost of a modest FER increase. Increasing $\lambda$ raises the stopping threshold, allowing more search and gradually narrowing the FER gap. At $\lambda=2048$, the FER remains close to DAI across all $E_b/N_0$ points while the average TEP count is still significantly lower.

\begin{figure}[t]
\centering
\begin{subfigure}{\columnwidth}
\centering
\begin{tikzpicture}
\begin{axis}[
    name=rmfer,
    width=\columnwidth,
    height=0.6\columnwidth,
    xmin=0, xmax=3,
    xtick={0,0.5,1,1.5,2,2.5,3},
    ymode=log,
    ymin=1e-5, ymax=1,
    xlabel={$E_b/N_0$ (dB)},
    ylabel={FER},
    grid=both,
    grid style={dotted,gray!50},
    tick label style={font=\scriptsize},
    label style={font=\small},
    legend style={font=\tiny, draw=none, fill=none, at={(0.02,0)}, anchor=south west},
    legend columns=1,
]
\addplot+[figcolor1, solid, line width=0.7pt, mark=o, mark size=1.8pt, mark options={solid, fill=white}] coordinates {(0,0.53333) (0.5,0.34483) (1,0.16835) (1.5,0.06832) (2,0.02433) (2.5,0.006617) (3,0.001654) (3.5,0.000316)};
\addlegendentry{NES, $\lambda=384$}
\addplot+[figcolor2, solid, line width=0.7pt, mark=diamond*, mark size=1.9pt, mark options={solid, fill=figcolor2}] coordinates {(0,0.51230) (0.5,0.33069) (1,0.15743) (1.5,0.06150) (2,0.02031) (2.5,0.005489) (3,0.001233) (3.5,0.000216)};
\addlegendentry{NES, $\lambda=1024$}
\addplot+[figcolor3, solid, line width=0.7pt, mark=triangle, mark size=1.9pt, mark options={solid, fill=white}] coordinates {(0,0.50075) (0.5,0.32072) (1,0.15342) (1.5,0.05842) (2,0.01908) (2.5,0.004952) (3,0.001067) (3.5,0.000170)};
\addlegendentry{NES, $\lambda=2048$}
\addplot+[figcolor10, solid, line width=0.7pt, mark=square*, mark size=1.8pt, mark options={solid, fill=figcolor10}] coordinates {(0,0.54675) (0.5,0.35336) (1,0.16946) (1.5,0.06589) (2,0.02052) (2.5,0.005000) (3,0.000870) (3.5,0.000107)};
\addlegendentry{DAI}
\addplot+[figcolor9, solid, line width=0.7pt, mark=triangle*, mark size=1.9pt, mark options={solid, fill=figcolor9}] coordinates {(0,0.54675) (0.5,0.35336) (1,0.16946) (1.5,0.06589) (2,0.02052) (2.5,0.005000) (3,0.000870) (3.5,0.000107)};
\addlegendentry{SAI}
\addplot+[gray, solid, line width=0.7pt, mark=star, mark size=1.9pt, mark options={solid, fill=gray}] coordinates {(0,0.48379) (0.5,0.30798) (1,0.14422) (1.5,0.05397) (2,0.01733) (2.5,0.00418) (3,0.00071) (3.5,0.00014)};
\addlegendentry{TSC}
\addplot+[figcolor6, solid, line width=0.7pt, mark=pentagon*, mark size=1.9pt, mark options={solid, fill=figcolor6}] coordinates {(0,0.31058) (0.5,0.24363) (1,0.16298) (1.5,0.10354) (2,0.06066) (2.5,0.03157) (3,0.01428) (3.5,0.00564)};
\addlegendentry{NES, $\lambda=12$}
\addplot+[figcolor10, solid, line width=0.7pt, mark=square*, mark size=1.8pt, mark options={solid, fill=figcolor10}, forget plot] coordinates {(0,0.32051) (0.5,0.25013) (1,0.16650) (1.5,0.10336) (2,0.05990) (2.5,0.03142) (3,0.01376) (3.5,0.00527)};
\addplot+[figcolor9, solid, line width=0.7pt, mark=triangle*, mark size=1.9pt, mark options={solid, fill=figcolor9}, forget plot] coordinates {(0,0.32051) (0.5,0.24988) (1,0.16496) (1.5,0.10312) (2,0.06047) (2.5,0.03134) (3,0.01370) (3.5,0.00533)};
\addplot+[gray, solid, line width=0.7pt, mark=star, mark size=1.9pt, mark options={solid, fill=gray}, forget plot] coordinates {(0,0.30817) (0.5,0.24021) (1,0.15929) (1.5,0.09994) (2,0.05718) (2.5,0.03027) (3,0.01316) (3.5,0.00515) (4,0.00176)};
\node[font=\scriptsize\bfseries\sffamily, color=navyblue] at (axis cs:2.8,0.006) {\textsf{\textbf{n\,=\,128}}};
\node[font=\scriptsize\bfseries\sffamily, color=navyblue] at (axis cs:2.8,0.05) {\textsf{\textbf{n\,=\,32}}};
\draw[black, thin] (axis cs:1.9,0.01) rectangle (axis cs:2.1,0.03);
\coordinate (rmzoom) at (axis cs:2.1,0.01);
\end{axis}
\begin{axis}[
    name=rminset,
    at={(rmfer.south east)}, anchor=south east,
    xshift=-45pt, yshift=15pt,
    width=0.32\columnwidth, height=0.32\columnwidth,
    xmin=1.9, xmax=2.1, xtick={2},
    ymode=log, ymin=0.015, ymax=0.03,
    ytick={0.015, 0.02, 0.03},
    yticklabels={$1.5\!\times\!10^{-2}$, $2\!\times\!10^{-2}$, $3\!\times\!10^{-2}$},
    grid=both, grid style={dotted,gray!50},
    tick label style={font=\tiny},
    axis background/.style={fill=white},
]
\addplot+[figcolor1, solid, line width=0.7pt, mark=o, mark size=1.4pt, mark options={solid, fill=white}] coordinates {(1.5,0.06832) (2,0.02433) (2.5,0.006617)};
\addplot+[figcolor2, solid, line width=0.7pt, mark=diamond*, mark size=1.4pt, mark options={solid, fill=figcolor2}] coordinates {(1.5,0.06150) (2,0.02031) (2.5,0.005489)};
\addplot+[figcolor3, solid, line width=0.7pt, mark=triangle, mark size=1.4pt, mark options={solid, fill=white}] coordinates {(1.5,0.05842) (2,0.01908) (2.5,0.004952)};
\addplot+[figcolor10, solid, line width=0.7pt, mark=square*, mark size=1.4pt, mark options={solid, fill=figcolor10}] coordinates {(1.5,0.06589) (2,0.02052) (2.5,0.005000)};
\addplot+[figcolor9, solid, line width=0.7pt, mark=triangle*, mark size=1.4pt, mark options={solid, fill=figcolor9}] coordinates {(1.5,0.06589) (2,0.02052) (2.5,0.005000)};
\addplot+[gray, solid, line width=0.7pt, mark=star, mark size=1.4pt, mark options={solid, fill=gray}] coordinates {(1.5,0.05397) (2,0.01733) (2.5,0.00418)};
\end{axis}
\draw[->, black, thin] (rmzoom) -- (rminset.north);
\end{tikzpicture}%
\vspace{-0.3cm}
\caption{Frame error rate.}
\label{fig:rm-fer}
\end{subfigure}\\[0.0cm]
\begin{subfigure}{\columnwidth}
\centering
\begin{tikzpicture}
\begin{axis}[
    width=\columnwidth,
    height=0.6\columnwidth,
    xmin=0, xmax=3,
    xtick={0,0.5,1,1.5,2,2.5,3},
    ymode=log,
    ymin=1, ymax=20000,
    xlabel={$E_b/N_0$ (dB)},
    ylabel={Average TEP},
    grid=both,
    grid style={dotted,gray!50},
    tick label style={font=\scriptsize},
    label style={font=\small},
    legend style={font=\tiny, draw=none, fill=none, at={(0.02,0)}, anchor=south west},
    legend columns=1,
]
\addplot+[figcolor1, solid, line width=0.7pt, mark=o, mark size=1.8pt, mark options={solid, fill=white}] coordinates {(0,323.2) (0.5,271.8) (1,186.5) (1.5,100.2) (2,42.2) (2.5,13.3) (3,3.9) (3.5,1.6)};
\addlegendentry{NES, $\lambda=384$}
\addplot+[figcolor2, solid, line width=0.7pt, mark=diamond*, mark size=1.9pt, mark options={solid, fill=figcolor2}] coordinates {(0,630.5) (0.5,532.6) (1,354.3) (1.5,180.0) (2,72.3) (2.5,20.9) (3,5.3) (3.5,1.8)};
\addlegendentry{NES, $\lambda=1024$}
\addplot+[figcolor3, solid, line width=0.7pt, mark=triangle, mark size=1.9pt, mark options={solid, fill=white}] coordinates {(0,1004.7) (0.5,830.7) (1,550.7) (1.5,276.4) (2,107.0) (2.5,29.2) (3,6.8) (3.5,2.0)};
\addlegendentry{NES, $\lambda=2048$}
\addplot+[figcolor10, solid, line width=0.7pt, mark=square*, mark size=1.8pt, mark options={solid, fill=figcolor10}] coordinates {(0,535.1) (0.5,582.3) (1,475.1) (1.5,312.9) (2,149.7) (2.5,55.3) (3,18.2) (3.5,6.3)};
\addlegendentry{DAI}
\addplot+[figcolor9, solid, line width=0.7pt, mark=triangle*, mark size=1.9pt, mark options={solid, fill=figcolor9}] coordinates {(0,535.0) (0.5,582.1) (1,474.8) (1.5,312.9) (2,149.7) (2.5,55.3) (3,18.2) (3.5,6.3)};
\addlegendentry{SAI}
\addplot+[gray, solid, line width=0.7pt, mark=star, mark size=1.9pt, mark options={solid, fill=gray}] coordinates {(0,15024.5) (0.5,13328.3) (1,10590.6) (1.5,7450.0) (2,4418.8) (2.5,2029.2) (3,726.1) (3.5,202.2)};
\addlegendentry{TSC}
\end{axis}
\end{tikzpicture}%
\vspace{-0.3cm}
\caption{Average tested error patterns.}
\label{fig:rm-tep}
\end{subfigure}
\caption{FER and average TEP versus $E_b/N_0$ for $\mathcal{C}_2[128,64]$ and $\mathcal{C}_2[32,16]$.}
\vspace{-0.1cm}
\label{fig:rm}
\end{figure}
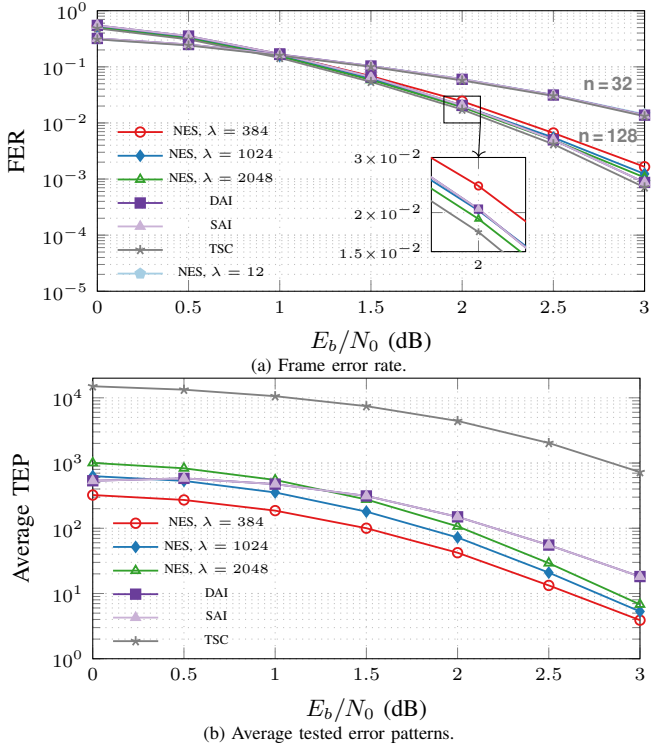

Fig. \ref{fig:rm} depicts the corresponding results for RM codes $\mathcal{C}_2[128,64]$ and $\mathcal{C}_2[32,16]$ with similar complexity--reliability trade-offs. At $\lambda=384$, the protocol reduces the average TEP count by a factor of $2\text{--}5\times$ relative to DAI, with a small FER penalty. At $\lambda=2048$, the FER closely tracks DAI and can even fall slightly below at certain $E_b/N_0$ points, while still offering meaningful complexity savings. Compared to eBCH, the TEP reductions relative to DAI are more conservative, though still far below TSC, while the FER remains closer to the baseline, reflecting the different decoding characteristics of the two code families.

\begin{figure}[t]
\centering
\begin{subfigure}{\columnwidth}
\centering
\begin{tikzpicture}
\begin{axis}[
    name=ldpcfer,
    width=\columnwidth,
    height=0.6\columnwidth,
    xmin=0, xmax=3,
    xtick={0,0.5,1,1.5,2,2.5,3},
    ymode=log,
    ymin=1e-5, ymax=1,
    xlabel={$E_b/N_0$ (dB)},
    ylabel={FER},
    grid=both,
    grid style={dotted,gray!50},
    tick label style={font=\scriptsize},
    label style={font=\small},
    legend style={font=\tiny, draw=none, fill=none, at={(0.02,0)}, anchor=south west},
    legend columns=1,
]
\addplot+[figcolor1, solid, line width=0.7pt, mark=o, mark size=1.8pt, mark options={solid, fill=white}] coordinates {(0,0.53937) (0.5,0.33344) (1,0.18054) (1.5,0.07250) (2,0.02473) (2.5,0.005827) (3,0.001405) (3.5,0.000289)};
\addlegendentry{NES, $\lambda=384$}
\addplot+[figcolor2, solid, line width=0.7pt, mark=diamond*, mark size=1.9pt, mark options={solid, fill=figcolor2}] coordinates {(0,0.52411) (0.5,0.32020) (1,0.16319) (1.5,0.06655) (2,0.02153) (2.5,0.004828) (3,0.001052) (3.5,0.000210)};
\addlegendentry{NES, $\lambda=1024$}
\addplot+[figcolor3, solid, line width=0.7pt, mark=triangle, mark size=1.9pt, mark options={solid, fill=white}] coordinates {(0,0.52219) (0.5,0.31250) (1,0.15957) (1.5,0.06380) (2,0.01969) (2.5,0.004223) (3,0.000902) (3.5,0.000179)};
\addlegendentry{NES, $\lambda=2048$}
\addplot+[figcolor10, solid, line width=0.7pt, mark=square*, mark size=1.8pt, mark options={solid, fill=figcolor10}] coordinates {(0,0.55463) (0.5,0.33841) (1,0.17088) (1.5,0.06564) (2,0.02044) (2.5,0.004116) (3,0.000771) (3.5,0.000129)};
\addlegendentry{DAI}
\addplot+[figcolor9, solid, line width=0.7pt, mark=triangle*, mark size=1.9pt, mark options={solid, fill=figcolor9}] coordinates {(0,0.55463) (0.5,0.33841) (1,0.17088) (1.5,0.06564) (2,0.02044) (2.5,0.004122) (3,0.000770) (3.5,0.000132)};
\addlegendentry{SAI}
\addplot+[gray, solid, line width=0.7pt, mark=star, mark size=1.9pt, mark options={solid, fill=gray}] coordinates {(0,0.50865) (0.5,0.29913) (1,0.14793) (1.5,0.05803) (2,0.01775) (2.5,0.00340) (3,0.00067) (3.5,0.00013)};
\addlegendentry{TSC}
\addplot+[figcolor6, solid, line width=0.7pt, mark=pentagon*, mark size=1.9pt, mark options={solid, fill=figcolor6}] coordinates {(0,0.36809) (0.5,0.26787) (1,0.19091) (1.5,0.13089) (2,0.07794) (2.5,0.04606) (3,0.02424) (3.5,0.01329)};
\addlegendentry{NES, $\lambda=12$}
\addplot+[figcolor10, solid, line width=0.7pt, mark=square*, mark size=1.8pt, mark options={solid, fill=figcolor10}, forget plot] coordinates {(0,0.37369) (0.5,0.26867) (1,0.19501) (1.5,0.12974) (2,0.07763) (2.5,0.04477) (3,0.02371) (3.5,0.01234)};
\addplot+[figcolor9, solid, line width=0.7pt, mark=triangle*, mark size=1.9pt, mark options={solid, fill=figcolor9}, forget plot] coordinates {(0,0.37369) (0.5,0.26867) (1,0.19516) (1.5,0.12985) (2,0.07768) (2.5,0.04489) (3,0.02393) (3.5,0.01233)};
\addplot+[gray, solid, line width=0.7pt, mark=star, mark size=1.9pt, mark options={solid, fill=gray}, forget plot] coordinates {(0,0.35663) (0.5,0.26462) (1,0.18730) (1.5,0.12523) (2,0.07468) (2.5,0.04449) (3,0.02352) (3.5,0.01164) (4,0.00484)};
\node[font=\scriptsize\bfseries\sffamily, color=navyblue] at (axis cs:2.8,0.006) {\textsf{\textbf{n\,=\,128}}};
\node[font=\scriptsize\bfseries\sffamily, color=navyblue] at (axis cs:2.8,0.055) {\textsf{\textbf{n\,=\,32}}};
\draw[black, thin] (axis cs:1.9,0.01) rectangle (axis cs:2.1,0.03);
\coordinate (ldpczoom) at (axis cs:2.1,0.01);
\end{axis}
\begin{axis}[
    name=ldpcinset,
    at={(ldpcfer.south east)}, anchor=south east,
    xshift=-45pt, yshift=15pt,
    width=0.32\columnwidth, height=0.32\columnwidth,
    xmin=1.9, xmax=2.1, xtick={2},
    ymode=log, ymin=0.015, ymax=0.03,
    ytick={0.015, 0.02, 0.03},
    yticklabels={$1.5\!\times\! 10^{-2}$, $2\!\times\!10^{-2}$, $3\!\times\!10^{-2}$},
    grid=both, grid style={dotted,gray!50},
    tick label style={font=\tiny},
    axis background/.style={fill=white},
]
\addplot+[figcolor1, solid, line width=0.7pt, mark=o, mark size=1.4pt, mark options={solid, fill=white}] coordinates {(1.5,0.07250) (2,0.02473) (2.5,0.005827)};
\addplot+[figcolor2, solid, line width=0.7pt, mark=diamond*, mark size=1.4pt, mark options={solid, fill=figcolor2}] coordinates {(1.5,0.06655) (2,0.02153) (2.5,0.004828)};
\addplot+[figcolor3, solid, line width=0.7pt, mark=triangle, mark size=1.4pt, mark options={solid, fill=white}] coordinates {(1.5,0.06380) (2,0.01969) (2.5,0.004223)};
\addplot+[figcolor10, solid, line width=0.7pt, mark=square*, mark size=1.4pt, mark options={solid, fill=figcolor10}] coordinates {(1.5,0.06564) (2,0.02044) (2.5,0.004116)};
\addplot+[figcolor9, solid, line width=0.7pt, mark=triangle*, mark size=1.4pt, mark options={solid, fill=figcolor9}] coordinates {(1.5,0.06564) (2,0.02044) (2.5,0.004122)};
\addplot+[gray, solid, line width=0.7pt, mark=star, mark size=1.4pt, mark options={solid, fill=gray}] coordinates {(1.5,0.05803) (2,0.01775) (2.5,0.00340)};
\end{axis}
\draw[->, black, thin] (ldpczoom) -- (ldpcinset.north);
\end{tikzpicture}%
\vspace{-0.3cm}
\caption{Frame error rate.}
\label{fig:ldpc-fer}
\end{subfigure}\\[0.0cm]
\begin{subfigure}{\columnwidth}
\centering
\begin{tikzpicture}
\begin{axis}[
    width=\columnwidth,
    height=0.6\columnwidth,
    xmin=0, xmax=3,
    xtick={0,0.5,1,1.5,2,2.5,3},
    ymode=log,
    ymin=1, ymax=20000,
    xlabel={$E_b/N_0$ (dB)},
    ylabel={Average TEP},
    grid=both,
    grid style={dotted,gray!50},
    tick label style={font=\scriptsize},
    label style={font=\small},
    legend style={font=\tiny, draw=none, fill=none, at={(0.02,0)}, anchor=south west},
    legend columns=1,
]
\addplot+[figcolor1, solid, line width=0.7pt, mark=o, mark size=1.8pt, mark options={solid, fill=white}] coordinates {(0,311.6) (0.5,260.3) (1,188.9) (1.5,106.6) (2,48.6) (2.5,18.2) (3,6.1) (3.5,2.3)};
\addlegendentry{NES, $\lambda=384$}
\addplot+[figcolor2, solid, line width=0.7pt, mark=diamond*, mark size=1.9pt, mark options={solid, fill=figcolor2}] coordinates {(0,615.7) (0.5,511.8) (1,354.0) (1.5,193.3) (2,82.9) (2.5,29.1) (3,8.7) (3.5,2.8)};
\addlegendentry{NES, $\lambda=1024$}
\addplot+[figcolor3, solid, line width=0.7pt, mark=triangle, mark size=1.9pt, mark options={solid, fill=white}] coordinates {(0,990.3) (0.5,794.2) (1,544.2) (1.5,294.1) (2,125.1) (2.5,41.5) (3,11.6) (3.5,3.5)};
\addlegendentry{NES, $\lambda=2048$}
\addplot+[figcolor10, solid, line width=0.7pt, mark=square*, mark size=1.8pt, mark options={solid, fill=figcolor10}] coordinates {(0,553.5) (0.5,596.9) (1,509.3) (1.5,311.1) (2,178.2) (2.5,75.1) (3,26.3) (3.5,8.9)};
\addlegendentry{DAI}
\addplot+[figcolor9, solid, line width=0.7pt, mark=triangle*, mark size=1.9pt, mark options={solid, fill=figcolor9}] coordinates {(0,553.3) (0.5,596.8) (1,509.1) (1.5,311.1) (2,178.2) (2.5,75.1) (3,26.3) (3.5,8.9)};
\addlegendentry{SAI}
\addplot+[gray, solid, line width=0.7pt, mark=star, mark size=1.9pt, mark options={solid, fill=gray}] coordinates {(0,15094.9) (0.5,13851.0) (1,11300.2) (1.5,8211.4) (2,5046.6) (2.5,2496.5) (3,966.8) (3.5,299.0)};
\addlegendentry{TSC}
\end{axis}
\end{tikzpicture}%
\vspace{-0.3cm}
\caption{Average tested error patterns.}
\label{fig:ldpc-tep}
\end{subfigure}
\caption{FER and average TEP versus $E_b/N_0$ for $\mathcal{C}_3[128,64]$ and $\mathcal{C}_3[32,16]$.}
\vspace{-0.1cm}
\label{fig:ldpc}
\end{figure}
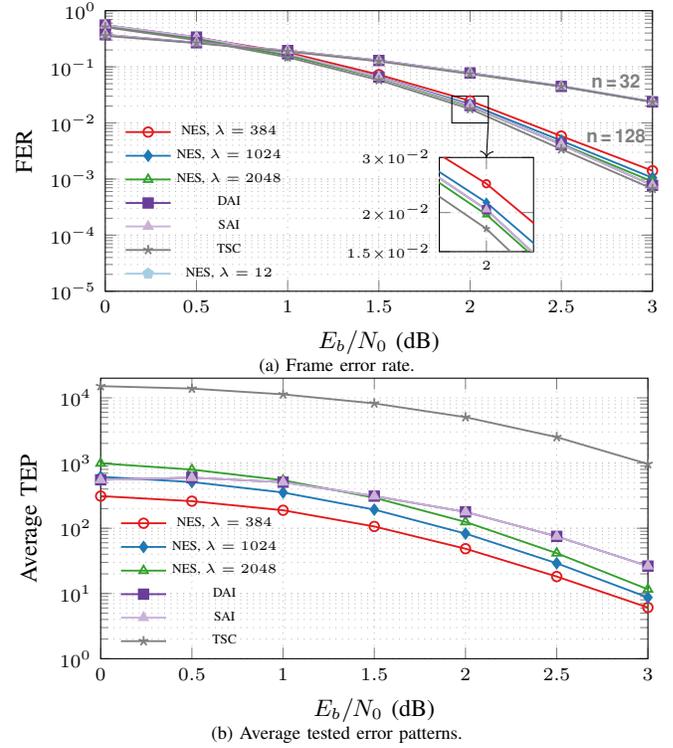

Fig. \ref{fig:ldpc} shows the results for LDPC codes $\mathcal{C}_3[128,64]$ and $\mathcal{C}_3[32,16]$. The proposed protocol follows the same trend, with TEP reductions comparable to RM but smaller than eBCH, while all operating points remain well below TSC in average TEP count. These results confirm that the NES protocol generalizes across different code families, requiring only retraining on the target code without any architectural or hyperparameter changes.

The $[32,16]$ results across all three code families confirm that NES also generalizes to shorter block lengths. The FER remains nearly identical to the baselines, though the TEP reductions are less observable than in the $[128,64]$ case due to the lower search complexity at shorter block lengths.

\vspace{-0.1cm}
\section{Conclusion}
We proposed a neural early stopping (NES) framework for LC-OSD that learns to predict if further search is likely to improve the decoded output, replacing hand-crafted methods with a data-driven protocol. NES provides tunable control over the complexity--reliability trade-off, and uses a compact representation of the decoder state to make informed stopping decisions. The resulting protocol generalizes across all $E_b/N_0$ operating points from a single trained model and transfers across code families without architectural changes, achieving significant TEP reductions with minimal FER degradation.

\bibliographystyle{IEEEtran}
\bibliography{ref}

@article{polyanskiy2010finite,
  author  = {Yury Polyanskiy and H. Vincent Poor and Sergio Verd{\'u}},
  title   = {Channel coding rate in the finite blocklength regime},
  journal = {IEEE Trans. Inf. Theory},
  year    = {2010},
  volume  = {56},
  number  = {5},
  pages   = {2307--2359},
  month   = may,
  doi     = {10.1109/TIT.2010.2043769}
}

@article{chen2018urllc,
  author  = {H. Chen and R. Abbas and P. Cheng and M. Shirvanimoghaddam and others},
  title   = {Ultra-reliable low latency cellular networks: Use cases, challenges and approaches},
  journal = {IEEE Commun. Mag.},
  year    = {2018},
  volume  = {56},
  number  = {12},
  pages   = {119--125},
  month   = dec
}

@article{shirvanimoghaddam2019short,
  author  = {M. Shirvanimoghaddam and M. S. Mohamadi and R. Abbas and others},
  title   = {Short block-length codes for ultra-reliable low-latency communications},
  journal = {IEEE Commun. Mag.},
  year    = {2019},
  volume  = {57},
  number  = {2},
  pages   = {130--137},
  month   = feb
}

@article{fossorier1995,
  author  = {Marc P. C. Fossorier and Shu Lin},
  title   = {Soft-decision decoding of linear block codes based on ordered statistics},
  journal = {IEEE Trans. Inf. Theory},
  year    = {1995},
  volume  = {41},
  number  = {5},
  pages   = {1379--1396},
  month   = sep
}

@article{yue2021pbosd,
  author  = {Chentao Yue and Mahyar Shirvanimoghaddam and Giyoon Park and Ok-Sun Park and Branka Vucetic and Yonghui Li},
  title   = {Probability-based ordered-statistics decoding for short block codes},
  journal = {IEEE Commun. Lett.},
  year    = {2021},
  volume  = {25},
  number  = {6},
  pages   = {1791--1795},
  month   = jun,
  doi     = {10.1109/LCOMM.2021.3058978}
}

@inproceedings{wang2021eosd,
  author    = {Fei Wang and Jian Jiao and Ke Zhang and Shaohua Wu and Yonghui Li and Qinyu Zhang},
  title     = {Efficient ordered statistics decoder for ultra-reliable low latency communications},
  booktitle = {Proc. IEEE Int. Conf. Commun. (ICC)},
  year      = {2021},
  doi       = {10.1109/ICC42927.2021.9500810}
}

@article{fossorier2002isr,
  author  = {Marc P. C. Fossorier},
  title   = {Reliability-based soft-decision decoding with iterative information set reduction},
  journal = {IEEE Trans. Inf. Theory},
  year    = {2002},
  volume  = {48},
  number  = {12},
  pages   = {3101--3106},
  month   = dec,
  doi     = {10.1109/TIT.2002.805089}
}

@article{jin2007bias,
  author  = {Wenyi Jin and Marc P. C. Fossorier},
  title   = {Reliability-based soft-decision decoding with multiple biases},
  journal = {IEEE Trans. Inf. Theory},
  year    = {2007},
  volume  = {53},
  number  = {1},
  pages   = {105--120},
  month   = jan,
  doi     = {10.1109/TIT.2006.887510}
}

@inproceedings{wang2022lcosd,
  author    = {Yiwen Wang and Jifan Liang and Xiao Ma},
  title     = {Local constraint-based ordered statistics decoding for short block codes},
  booktitle = {Proc. IEEE Inf. Theory Workshop (ITW)},
  year      = {2022},
  doi       = {10.1109/ITW54588.2022.9965916}
}

@article{liang2023lcosd,
  author    = {Jifan Liang and Yiwen Wang and Suihua Cai and Xiao Ma},
  title     = {A Low-Complexity Ordered Statistic Decoding of Short Block Codes},
  journal   = {IEEE Commun. Lett.},
  volume    = {27},
  number    = {2},
  pages     = {400--403},
  month     = feb,
  year      = {2023},
  doi       = {10.1109/LCOMM.2022.3222819}
}

@article{yue2022leosd,
  author  = {Chentao Yue and Mahyar Shirvanimoghaddam and Giyoon Park and Ok-Sun Park and Branka Vucetic and Yonghui Li},
  title   = {Linear-equation ordered-statistics decoding},
  journal = {IEEE Trans. Commun.},
  year    = {2022},
  volume  = {70},
  number  = {11},
  pages   = {7105--7123},
  month   = nov,
  doi     = {10.1109/TCOMM.2022.3207206}
}

@article{choi2021fast,
  author  = {Changryoul Choi and Jechang Jeong},
  title   = {Fast soft decision decoding algorithm for linear block codes using permuted generator matrices},
  journal = {IEEE Commun. Lett.},
  year    = {2021},
  volume  = {25},
  number  = {12},
  pages   = {3775--3779},
  month   = dec,
  doi     = {10.1109/LCOMM.2021.3097322}
}

@article{choi2019fastscalable,
  author  = {Changryoul Choi and Jechang Jeong},
  title   = {Fast and scalable soft decision decoding of linear block codes},
  journal = {IEEE Commun. Lett.},
  year    = {2019},
  volume  = {23},
  number  = {10},
  pages   = {1753--1756},
  month   = oct,
  doi     = {10.1109/LCOMM.2019.2927218}
}

@inproceedings{yue2022age,
  author    = {Chentao Yue and Mahyar Shirvanimoghaddam and Branka Vucetic and Yonghui Li},
  title     = {Ordered-statistics decoding with adaptive Gaussian elimination reduction for short codes},
  booktitle = {Proc. IEEE Globecom Workshops},
  year      = {2022},
  doi       = {10.1109/GCWkshps56602.2022.10008654}
}

@article{li2024ibu,
  author  = {Xihao Li and Wenhao Chen and Li Chen and Huazi Zhang},
  title   = {Iterative basis update for ordered statistics decoding of linear block codes},
  journal = {IEEE Commun. Lett.},
  year    = {2024},
  volume  = {28},
  number  = {9},
  pages   = {1981--1985},
  month   = sep,
  doi     = {10.1109/LCOMM.2024.3425388}
}

@inproceedings{yang2022lowlatency,
  author    = {Lijia Yang and Li Chen},
  title     = {Low-latency ordered statistics decoding of {BCH} codes},
  booktitle = {Proc. IEEE Inf. Theory Workshop (ITW)},
  year      = {2022},
  doi       = {10.1109/ITW54588.2022.9965799}
}

@article{li2024boosting,
  author  = {Guangwen Li and Xiao Yu},
  title   = {Boosting ordered statistics decoding of short {LDPC} codes with simple neural network models},
  journal = {IEEE Commun. Lett.},
  year    = {2024},
  volume  = {28},
  number  = {12},
  pages   = {2714--2718},
  month   = dec,
  doi     = {10.1109/LCOMM.2024.3475874}
}

@inproceedings{yue2025guesswork,
  author    = {Chentao Yue and Branka Vucetic and Yonghui Li},
  title     = {Guesswork complexity of ordered statistics decoding and its saturation threshold},
  booktitle = {Proc. IEEE Int. Symp. Inf. Theory (ISIT)},
  year      = {2025},
  doi       = {10.1109/ISIT63088.2025.11195325}
}

\end{document}